\documentclass[aps,nopacs,amsmath,amssymb,twocolumn]{revtex4}
\usepackage{graphicx}

\pdfoutput=1 

\def\beq{\begin{eqnarray}}
\def\eeq{\end{eqnarray}}
\def\be{\begin{equation}}
\def\ee{\end{equation}}
\def\eq{&=&}

\def\bm{\begin{math}}
\def\me{\end{math}}

\def\lb{\label}
\def\q{\quad}
\def\qq{\qquad}
\newcommand \nn {\nonumber}
\newcommand \bei {\begin{itemize}}
\newcommand \eei  {\end{itemize}}

\newcommand \nt   {\nonumber \\ }

\begin{document}
\bibliographystyle{apsrev}

\title{Oscillations in aggregation-shattering processes}
\author{S. A. Matveev$^{1,2,3}$}
\author{P. L. Krapivsky$^{4}$}
\author{A. P. Smirnov$^{2,3}$}
\author{E. E. Tyrtyshnikov$^{2,3}$}
\author{N. V. Brilliantov$^{5}$}
\affiliation{$^{1}$Skolkovo Institute of Science and Technology, Moscow, Russia}
\affiliation{$^{2}$Faculty of Computational Mathematics and Cybernetics, Lomonosov MSU, Moscow Russia }
\affiliation{$^{3}$Institute of Numerical Mathematics RAS, Moscow Russia }
\affiliation{$^{4}$Department of Physics, Boston University, Boston, MA 02215, USA}
\affiliation{$^{5}$Department of Mathematics, University of Leicester, Leicester LE1 7RH, United Kingdom }

\date{\today}

\begin{abstract}
We observe never-ending oscillations in systems undergoing  aggregation and collision-controlled shattering. Specifically, we investigate aggregation-shattering processes with aggregation kernels $K_{i,j} = (i/j)^a+(j/i)^a$ and shattering kernels $F_{i,j}=\lambda K_{i,j}$, where $i$ and $j$ are cluster sizes and parameter $\lambda$ quantifies the strength of shattering. When $0\leq a<1/2$, there are no oscillations and the system monotonically approaches to a steady state for all values of $\lambda$; in this region we obtain an analytical solution for the stationary cluster size distribution. Numerical solutions of the
rate equations show that oscillations emerge in the $1/2 < a \leq 1$ range. When $\lambda$ is sufficiently large oscillations decay and eventually disappear, while for $\lambda < \lambda_c(a)$ oscillations apparently persist forever. Thus never-ending oscillations can arise in {\em closed} aggregation-shattering processes without sinks and sources of particles.

\end{abstract}

\maketitle

Two complimentary processes, aggregation and fragmentation \cite{Flory,Krapivsky,Leyvraz}, occur in numerous systems that dramatically differ in their spatial and temporal  scales. Reversible polymerization in solutions \cite{Flory} and merging of
prions (cell proteins) \cite{prions} are typical examples on the molecular scale. On somewhat larger scales
airborne particles perform Brownian motion in atmosphere and coalesce giving rise to smog droplets
\cite{Srivastava1982}. Aggregation of users in the Internet leads to the emergence of communities and forums
\cite{Krapivsky,Dorogov} which can further merge or split. Vortexes in a fluid flow merge and decompose
forming turbulent cascades \cite{Turbulence}. On much larger scales, aggregation-fragmentation processes take
place in planetary rings, like Saturn rings, where the particle size distribution is determined by a subtle
balance between aggregation and fragmentation of the rings particles \cite{PNAS,
stadnichuk2015smoluchowski,Cuzzi,Brill2009,Esposito_book}.

In spatially homogeneous well-mixed systems, aggregation and fragmentation processes are described by an infinite set of nonlinear ordinary differential equations (ODEs) for the concentrations of clusters of various masses.  Such equations are intractable apart from a few special cases. The long-time behavior, however, is occasionally known---the processes of aggregation and fragmentation act in the opposite directions and hence the cluster size distribution often becomes stationary in the long time limit~\footnote{There are a few exceptions when the typical cluster size diverges and/or the
system undergoes a non-equilibrium phase transition, see e.g. \cite{bk08}.}. The emergence of the stationary
cluster size distribution can be mathematically interpreted as the manifestation of the fixed point  of the
differential equations \cite{Strogatz}. For a single differential equation, fixed points determine the long
time behavior, while for two coupled differential equations the asymptotic behavior may be determined by a
fixed point or a limit cycle. In the case of infinitely many coupled ODEs, limit cycles are feasible,
yet they haven't been observed in aggregation-fragmentation processes. More precisely, there were signs of oscillations in a few open systems usually driven by constant source of monomers and by sink of large clusters. In this paper we report oscillations in a {\em closed} system undergoing aggregation and collision-controlled fragmentation.

\begin{figure}[h]
\begin{center}
\includegraphics[scale=0.1]{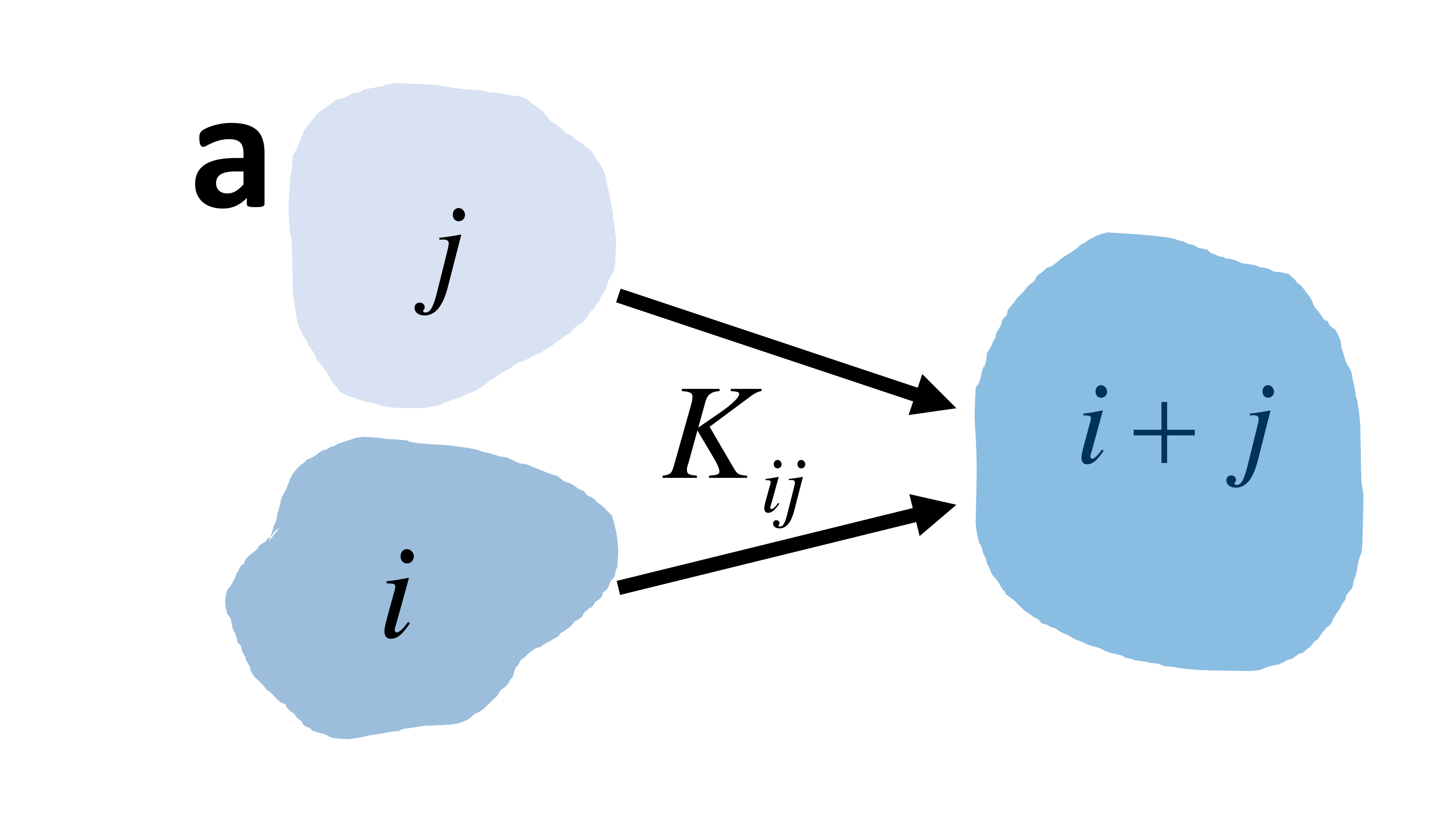} \qq \qq
\includegraphics[scale=0.1]{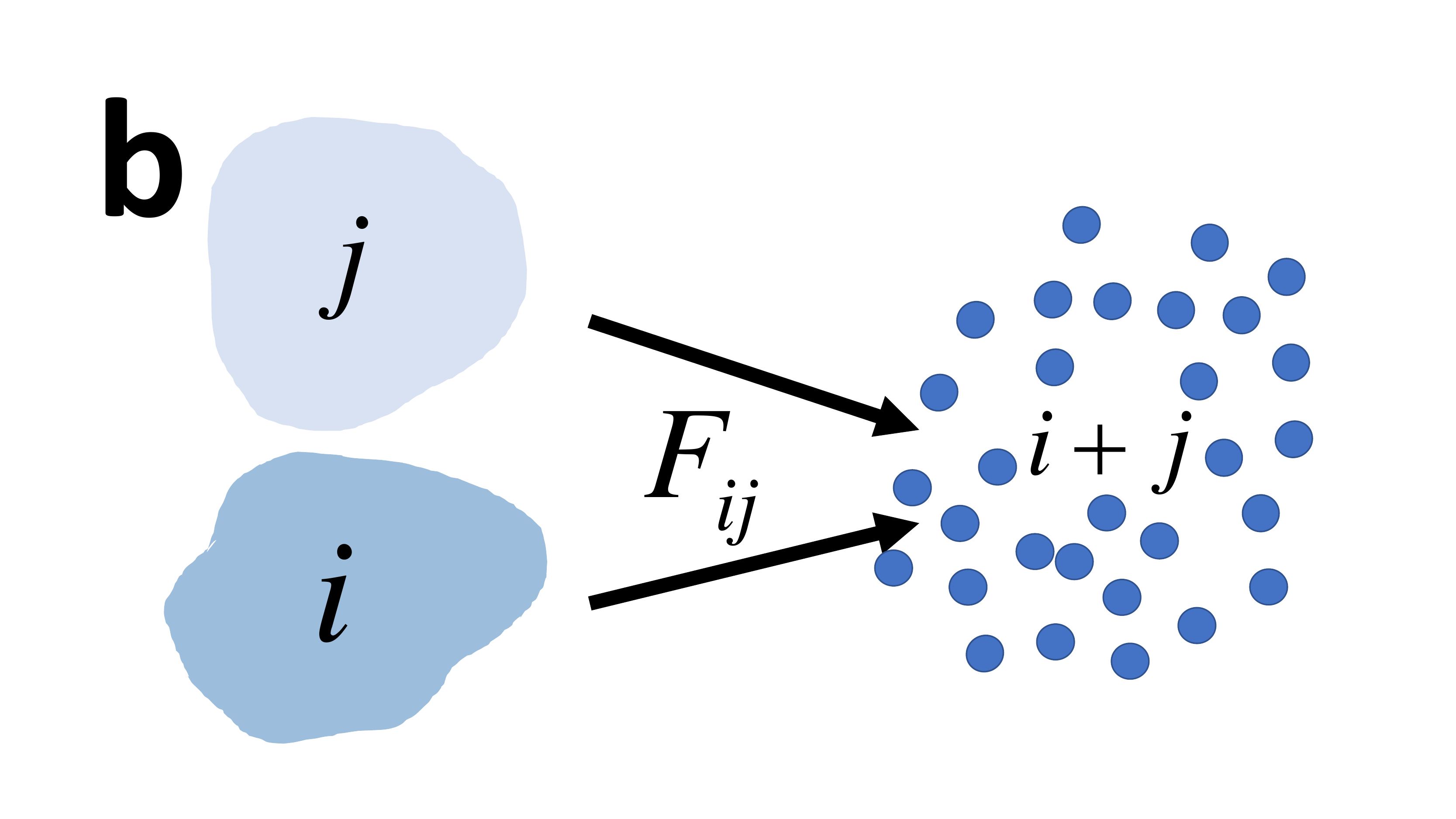}
\end{center}
\caption{Aggregation (a) and shattering (b) of clusters.} \label{fig:Agg_Frag}
\end{figure}

In the most important case of binary aggregation the collision between two clusters comprising
 $i$ and $j$ monomers may result in the formation of a joint aggregate of $i+j$ monomers. Symbolically, $[i] +[j]\xrightarrow{K_{i,j}} [i+j], $ where $K_{ij}$ is the merging rate (see Fig.~\ref{fig:Agg_Frag}). Let $n_k$ be the concentration of
clusters that  contain $k$ monomers. These quantities obey the Smoluchowski equations
\cite{Krapivsky,Leyvraz}:

\be \lb{eq:Smol1} \frac{dn_k}{dt} = \frac{1}{2}\sum_{i+j=k}K_{i,j}n_i n_j - n_k \sum_{i=1}^{\infty}K_{i,k}n_i
.  \ee
The first gain term on the right-hand side gives the formation rate  of $k$-mers from smaller clusters, while
the second terms describes the disappearance of $k$-mers due to collisions with other clusters (the factor
$1/2$ prevents double counting).

\vspace{-0.2cm}

In this article we consider collision-controlled fragmentation, which is thought to be
responsible e.g. for interstellar dust clouds and planetary rings~\cite{KrapivskyBenNaim2003,Brill2009,PNAS,
stadnichuk2015smoluchowski}. We explore the extreme version, namely a complete shattering of two
colliding partners into monomers (see Fig.~\ref{fig:Agg_Frag}). Symbolically
$[i] +[j] \xrightarrow{F_{i,j}} \underbrace{[1]+[1]+\ldots [1]}_{i+j}$ where $F_{i,j}$ quantifies the
shattering rate. It has been shown \cite{PNAS} that this shattering model is rather
generic---realistic impact models with a strong dominance of small debris over the large ones yield the
same resulting cluster size distribution $n_k$. Following~\cite{PNAS}, we assume that the shattering and
aggregation kernels are proportional,

\be \lb{eq:CijAij}  F_{i,j} = \lambda K_{i,j}.  \ee
The parameter $\lambda$ characterizes the relative frequency of collisions leading to merging and
shattering.

Incorporating the shattering process with the shattering kernel~\eqref{eq:CijAij} into
Eqs.~\eqref{eq:Smol1} we arrive at

\beq
\label{eq:Model}
\frac{dn_1}{dt} \eq  -n_1 \sum\limits_{i=1}^{\infty} K_{1,i} n_i + \frac{\lambda}{2}
\sum\limits_{i = 2}^{\infty} \sum\limits_{j=2}^{\infty} (i+j) K_{i,j} n_i~ n_j \nt
 &+& \lambda n_1
\sum\limits_{j=2}^{\infty} j K_{1,j} n_j  \\
\frac{dn_k}{dt} \eq \frac{1}{2} \sum\limits_{i=1}^{k-1} K_{i,k-i} n_i n_{k-i} - (1 + \lambda) n_k
\sum\limits_{i=1}^{\infty} K_{k,i} n_i. \nn \eeq
Shattering leads to the increase of monomers explaining the gain terms in the first equation \eqref{eq:Model} and it leads to the decrease of the density of other clusters explaining the loss term in the second equation.

A microscopic analysis is needed to establish how the kernels $K_{i,j}$ and  $F_{i,j}$ depend on the masses, see e.g. 
\cite{PNAS,Brill2009}. The kernels are always symmetric, $K_{i,j}=K_{j,i}$, and in most applications homogeneous functions of $i$ and $j$. Aggregation-shattering equations \eqref{eq:Model} for the generalized product kernels, $K_{i,j} = (ij)^{\mu}$, have been investigated in \cite{PNAS}. A more general family of kernels, $ K_{i,j} =  i^{\nu}j^{\mu} + i^{\mu}j^{\nu}$, is often used in studies of aggregation, see e.g.  Ref.~\cite{connaughton2016universality} where a source of monomers and sink of large clusters was present. We shall focus on a special case of $\mu=-\nu$,

\be \lb{eq:Cija} K_{i,j} = i^{a}j^{-a} + i^{-a}j^{a} ,\ee
which is also  known as a generalized Brownian kernel \cite{ColmPaulJCP2012}. Below we always assume that
$a\leq 1$, since aggregation equations with kernel \eqref{eq:Cija} satisfying $a>1$ become ill-defined due to
instantaneous gelation \cite{Ernst1983,van87,bk,Laurencot1999,Malyushkin2001,Colm2011}.

Time-dependent analytical solutions of Eqs.~\eqref{eq:Model} have been found \cite{PNAS} only for the
simplest case of a constant kernel ($a=0$). The steady-state solutions have been obtained for a wider class
of models, including irreversible aggregation model with a monomer source~\cite{hayakawa1987irreversible},
aggregation-fragmentation model with the generalized product kernel \cite{PNAS} and for a somewhat similar
open system with a source of monomers and collisional evaporation of clusters with the kernel $ K_{i,j} =
i^{\nu}j^{\mu} + i^{\mu}j^{\nu}$~\cite{connaughton2016universality}. An open aggregating system with the same
coagulation kernel driven by input of monomers and supplemented with removal of large clusters has been
studied in \cite{Colm}. Stable oscillations have been numerically observed \cite{Colm} in this system   with
finite number of cluster species.  For a closed system consisting of monomers, dimers, trimers and exited
monomers, stable oscillations of concentrations have been reported \cite{Gorban}. Steady chemical
oscillations have been also found in a simple dimerization model (see e.g. \cite{dimer} and references
therein).

Here we consider  {\em closed}  systems undergoing aggregation and shattering processes, so there are no
sources and sinks of monomers and clusters. The aggregation and shattering kernels are described by
Eqs.~\eqref{eq:Cija} and \eqref{eq:CijAij}. One expects that in the closed system with two opposite processes
and without sinks and sources, a steady state is achieved. This is indeed the case when the parameter
$a<1/2$. Surprisingly, for $1/2<a \leq  1$ and small values of $\lambda$, a steady state is not reached and
instead cluster concentrations undergo never-ending oscillations.

An important advantage of the kernel \eqref{eq:Cija} is the possibility to apply highly efficient numerical methods. Here we exploit a fast and accurate method of time-integration of Smoluchowski-type kinetic equations developed in recent studies \cite{matveev2015fast,matveev2016fast, matveev2016tensor, Chaudhury2014, hackbusch2006efficient, hackbusch2007approximation}, that has been adopted for the discrete distribution of cluster sizes. In our simulations we use up to $N_{\rm eq} = 2^{19} \equiv 524,288$ equations; in practice, we choose $N_{\rm eq}$ in such a way, that the further
increase of $N_{\rm eq}$ does not impact the simulation results for $n_k$ within the numerical
accuracy~\footnote{In the Supplementary Material (SM) we justify that an infinite system \eqref{eq:Model}
with the kernel \eqref{eq:Cija} may be approximated with any requested accuracy by a finite number of
equations.}.

\begin{figure}[h!]
\begin{center}
\includegraphics[scale=0.2]{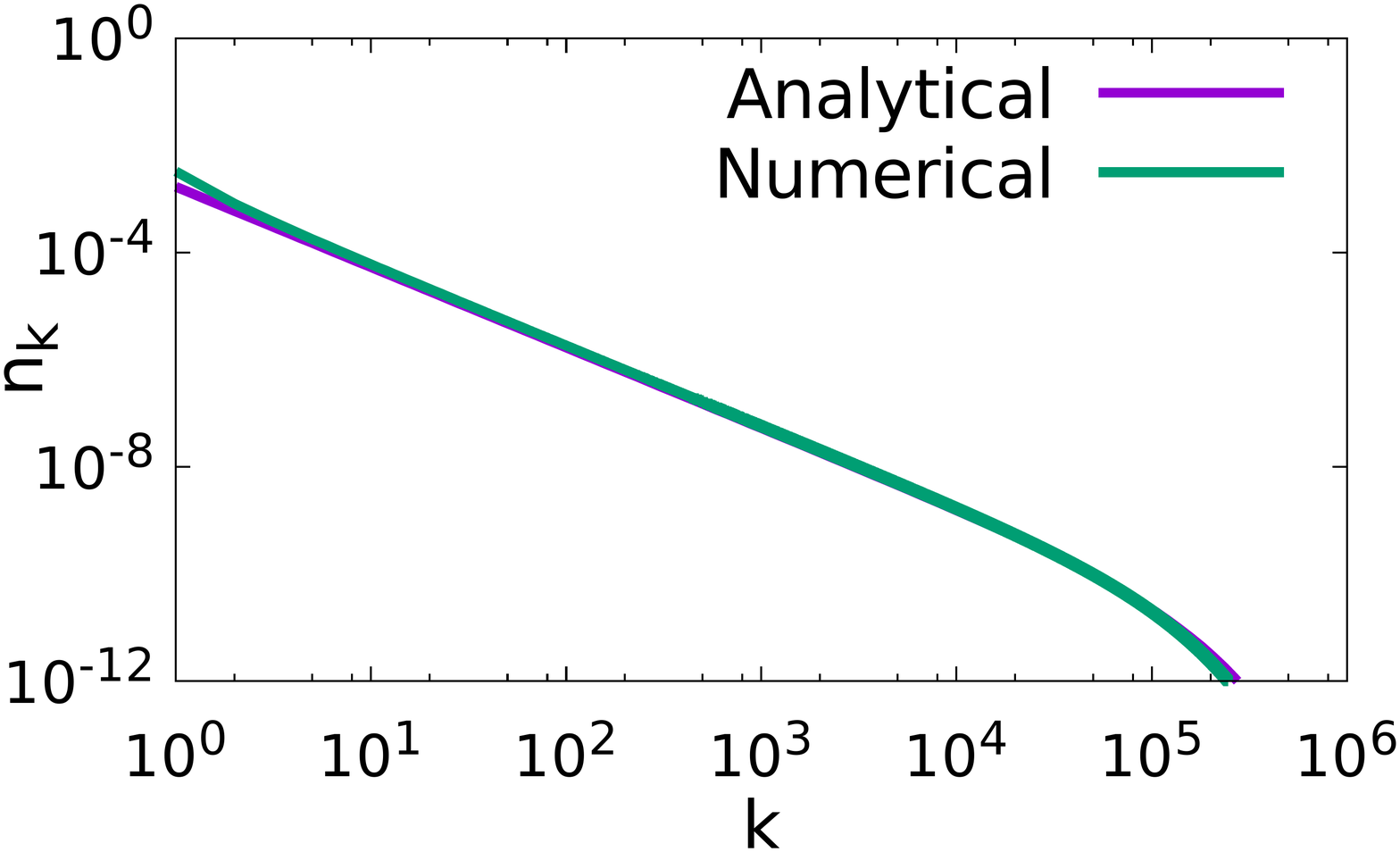} \qq
\includegraphics[scale=0.2]{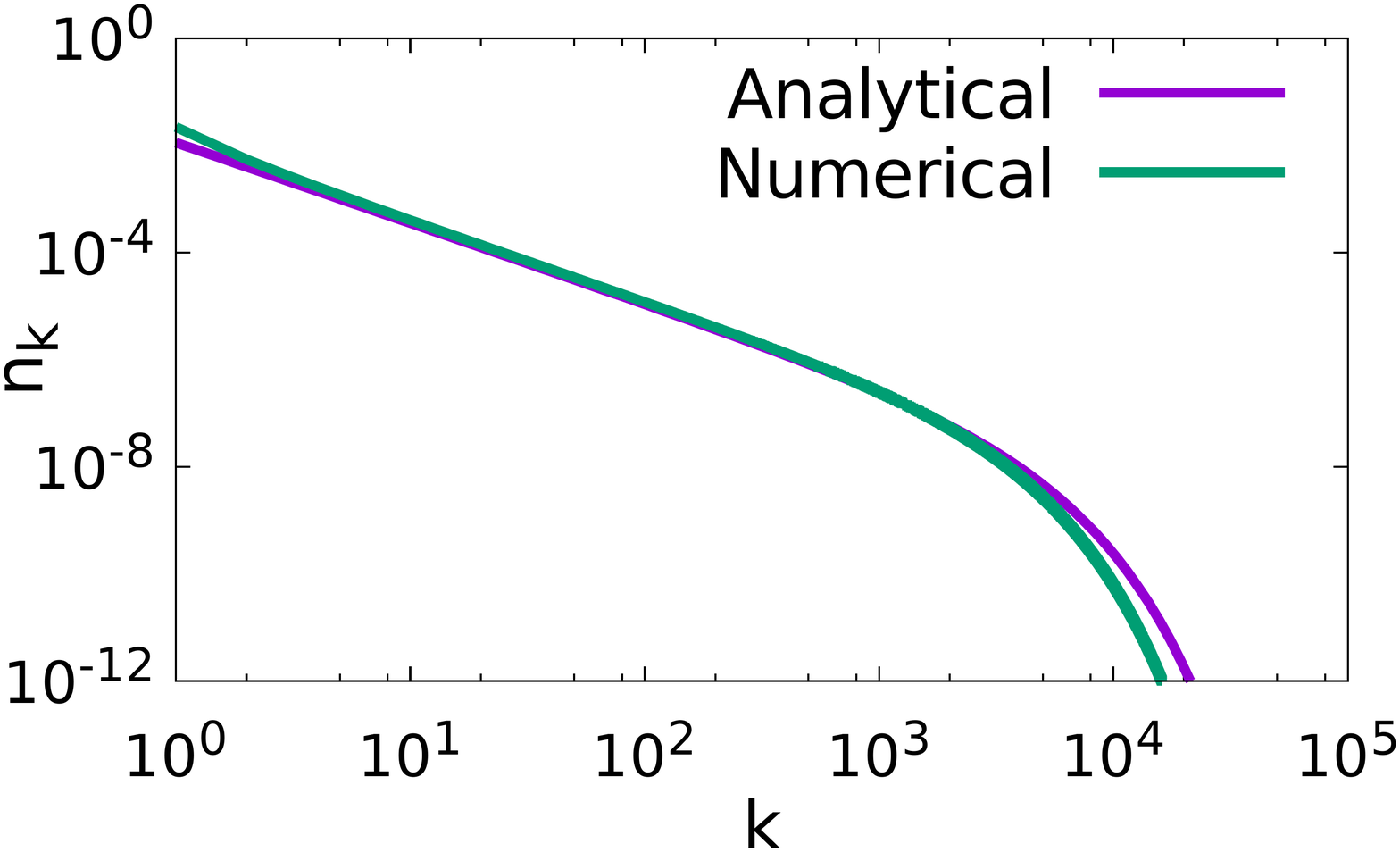}
\caption{Comparison of the steady-state numerical solution of Eqs.~\eqref{eq:Model} for the kernel
\eqref{eq:Cija} with $a=0.05$, $\lambda =0.003$  (top) and $a=0.1$, $\lambda=0.02$   (bottom) with the
analytical steady-state solution, Eq.~\eqref{eq:disfun2}. } \label{pic:st_a02}
\end{center}
\end{figure}

We confirm the efficiency and accuracy of the above fast-integration method for constant kernels ($a=0$),
comparing the numerical results with the available analytical solutions~\cite{PNAS} and find that the smaller
the parameter $\lambda$ the longer it takes for the system to reach the steady state, see the Supplementary
Material (SM). This tendency persists for the kernels \eqref{eq:Cija} with $a>0$.  We also observe that for
$a<1/2$ the system arrives at its steady-state with a monotonic evolution of the concentrations $n_k(t)$.
Moreover, the steady-state distribution of the cluster concentrations agrees fairly well with analytical
results for $n_k$ derived below. Figure \ref{pic:st_a02} illustrates the numerical solution for steady-state
distribution for $a=0.05$ and $a=0.1$ and different values of the shattering parameter $\lambda$. In the
language of dynamical systems \cite{Strogatz} we conclude that the system possesses a stable fixed point with
the steady-state cluster size distribution.

For $1/2< a\leq 1$, we also observed the relaxation to a steady state for sufficiently large $\lambda$, the relaxation occurs through oscillations for smaller $\lambda$, and when $\lambda<\lambda_c(a)$ the oscillations persist. We detected oscillations independently of the initial conditions.

The dynamic of the system described by Eqs.~\eqref{eq:Model} and \eqref{eq:Cija} is invariant with respect to re-scaling of the total mass density $M= \sum_{k=1}^{\infty} k n_k$ (see the SM). Below we report simulation results for the stepwise initial distribution

\be \lb{inicon}
n_k(0)=
\begin{cases}
0.1 & k=1,  \ldots, 10\\
0    & k>10
\end{cases}
\ee
and we also simulated the evolution starting with mono-disperse initial condition, $n_k(0) =M\delta_{1,k}$, with the same mass $M=5.5$. Unless explicitly stated, the results below correspond to the initial condition \eqref{inicon}.

\begin{figure}[ht]
\begin{center}
\includegraphics[scale=0.2]{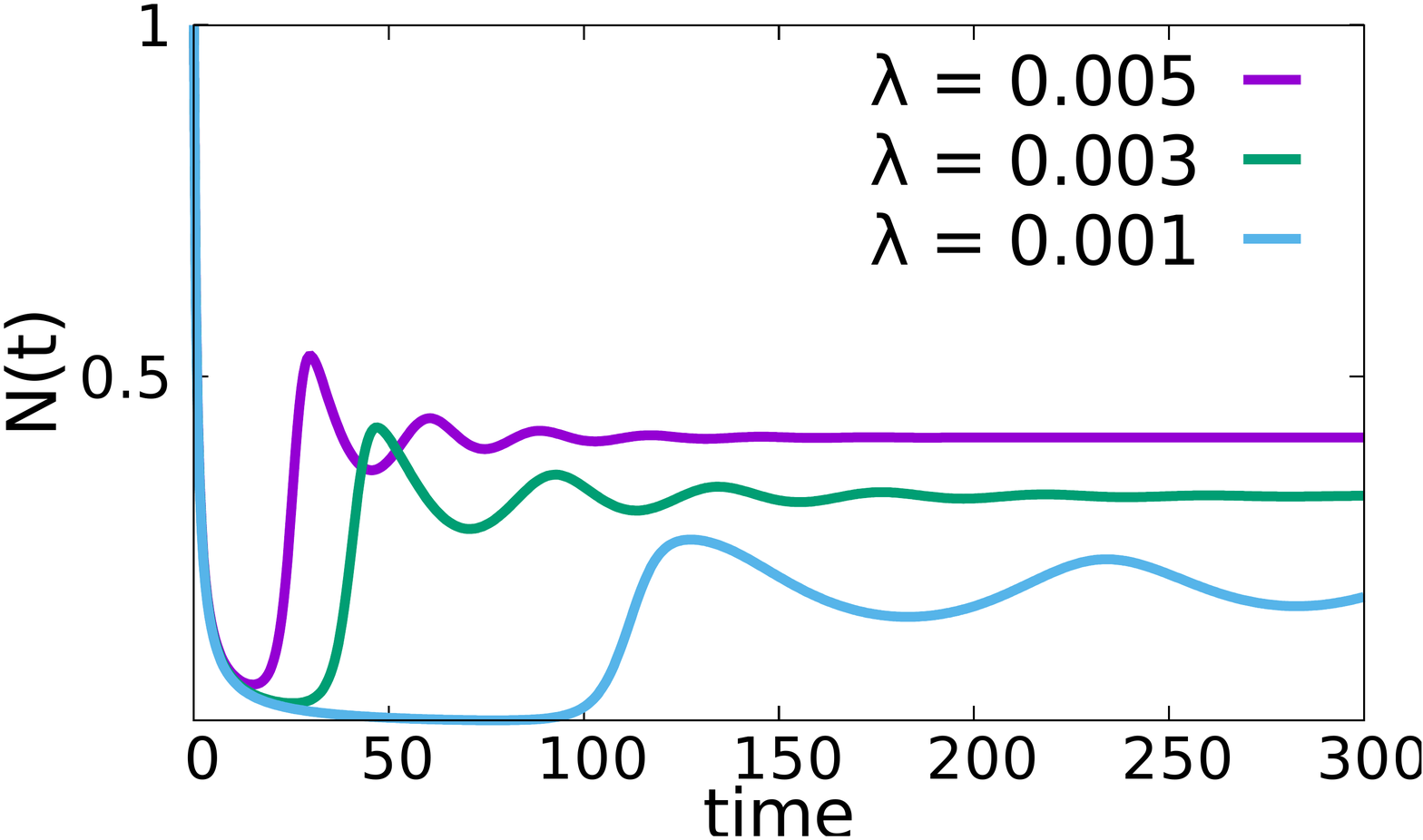}\\
~\\
\includegraphics[scale=0.2]{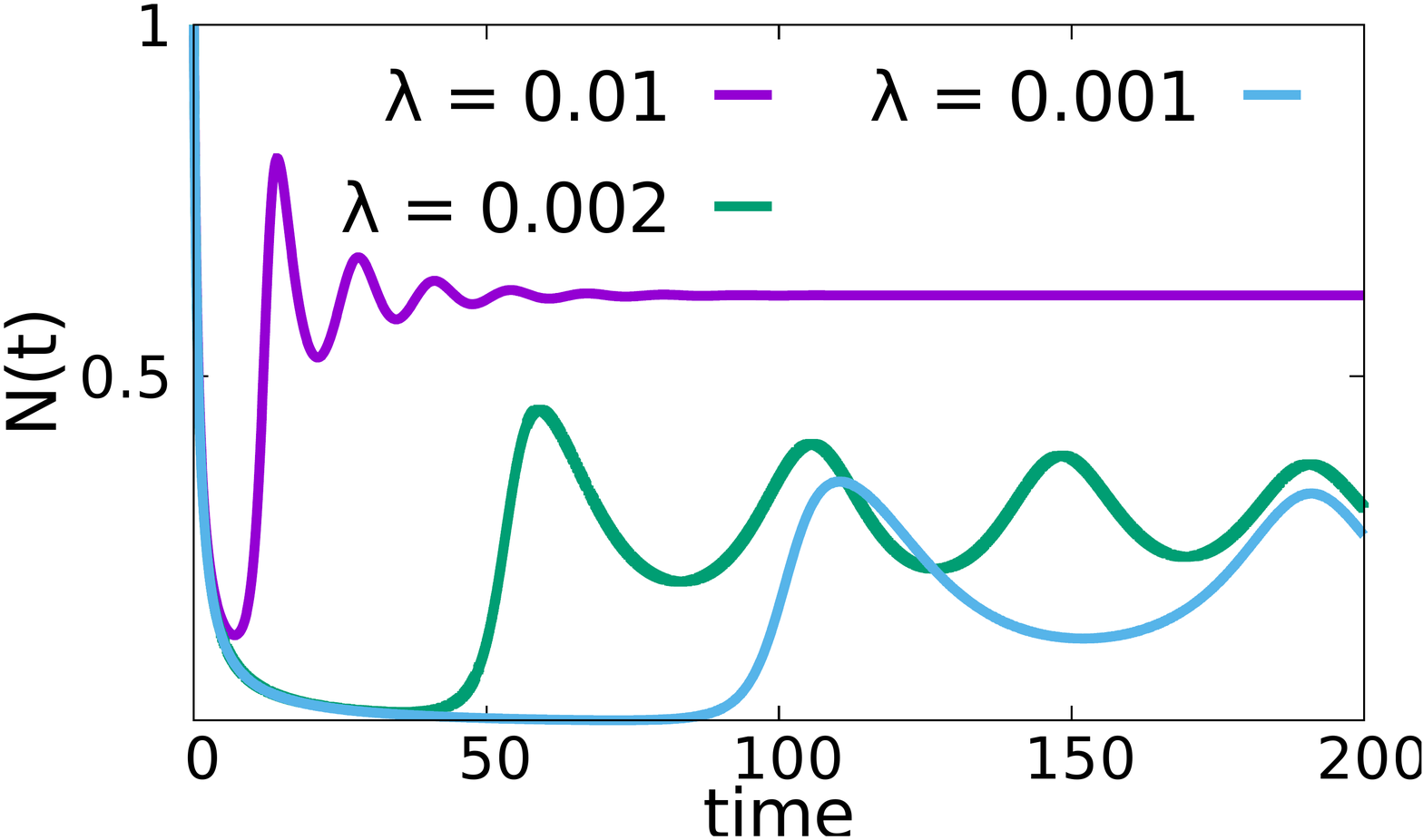} \qq
\end{center}
\caption{Time dependence of the clusters density, $N(t)$, for $a=0.7$ (top) and $a=0.75$ (bottom) and
different $\lambda$. For all these systems damped oscillations are found that tend to a steady-state. The
oscillations become more pronounced with increasing $a$ and decreasing $\lambda$.  }
\label{pic:Damping_osillations}
\end{figure}

\begin{figure}[h!]
\begin{center}
\includegraphics[scale=0.23,page=1]{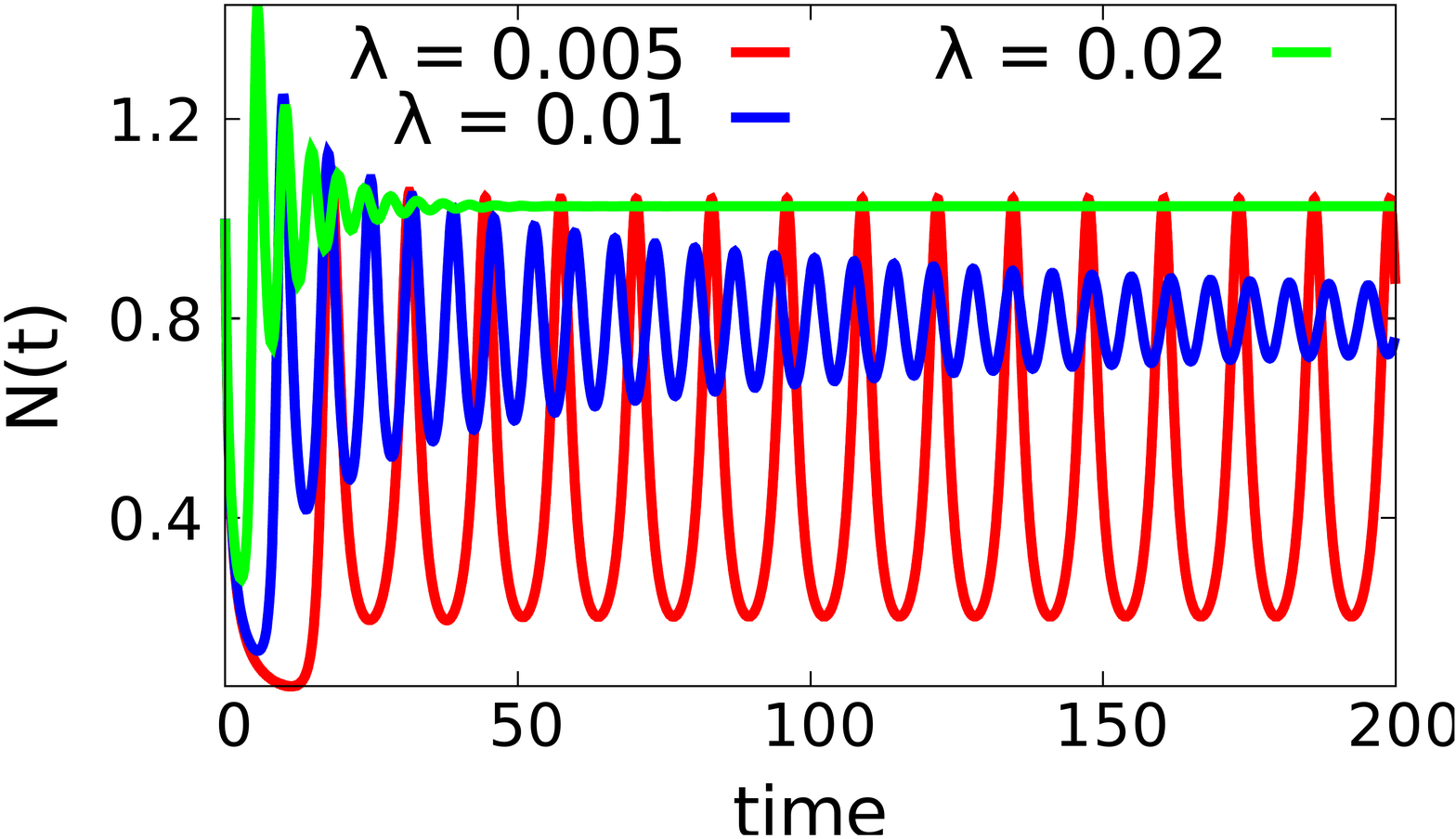} 
\end{center}
\caption{Time dependence of the cluster density, $N(t)$ for $a = 0.9$ and different $\lambda$. For small
$\lambda < \lambda_c(a)$ stable oscillations emerge. }
\label{pic:Osillations_integral}
\end{figure}

In Figs.~\ref{pic:Damping_osillations}--\ref{pic:Osillations_integral} we demonstrate the
time dependence of the cluster density $N(t)  = \sum_{k=1}^{\infty} n_k (t)$. Figure \ref{pic:Damping_osillations} shows that in the range $0.6 \leq a \leq 0.8$ and $0.001 \leq \lambda \leq 0.01$, the oscillations become more pronounced when $a$ increases and $\lambda$ decreases.

\begin{figure}[h!]
\begin{center}
\includegraphics[scale=0.5]{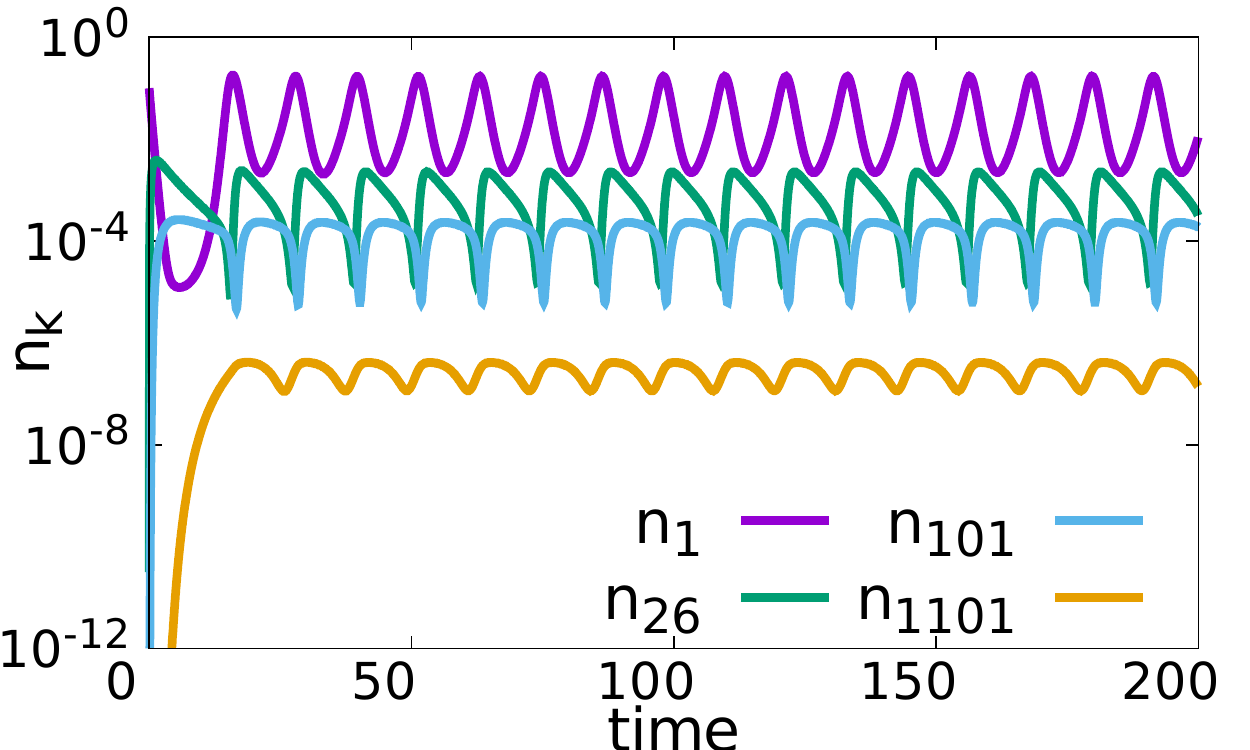} \q
\includegraphics[scale=0.5]{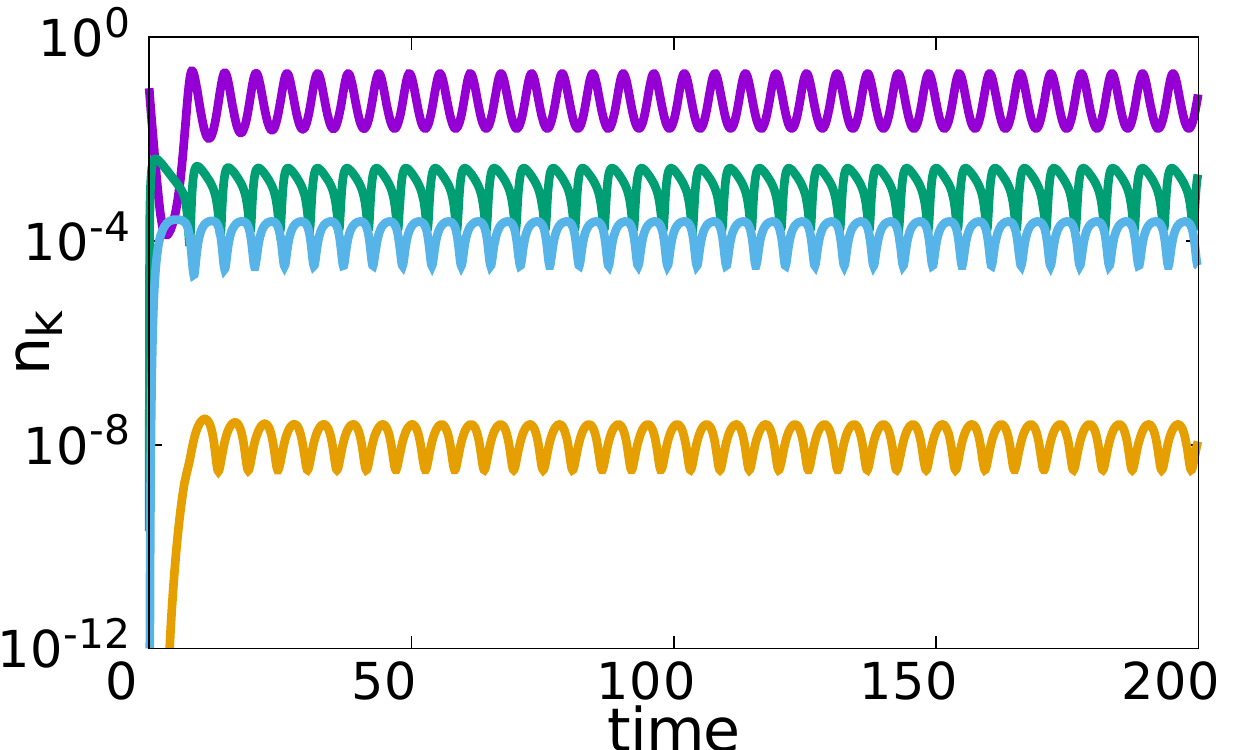}
\caption{Stationary oscillations of the aggregate concentrations for $a=0.95$ (top) and $a=1$ (bottom) and
$\lambda < \lambda_c(a)$. The shape of the oscillations depends on the aggregates' size;  the amplitude of
the oscillations decreases with the size. }
\label{pic:Oscil_095098}
\end{center}
\end{figure}

\begin{figure}[ht]
\begin{center}
\includegraphics[scale=0.23]{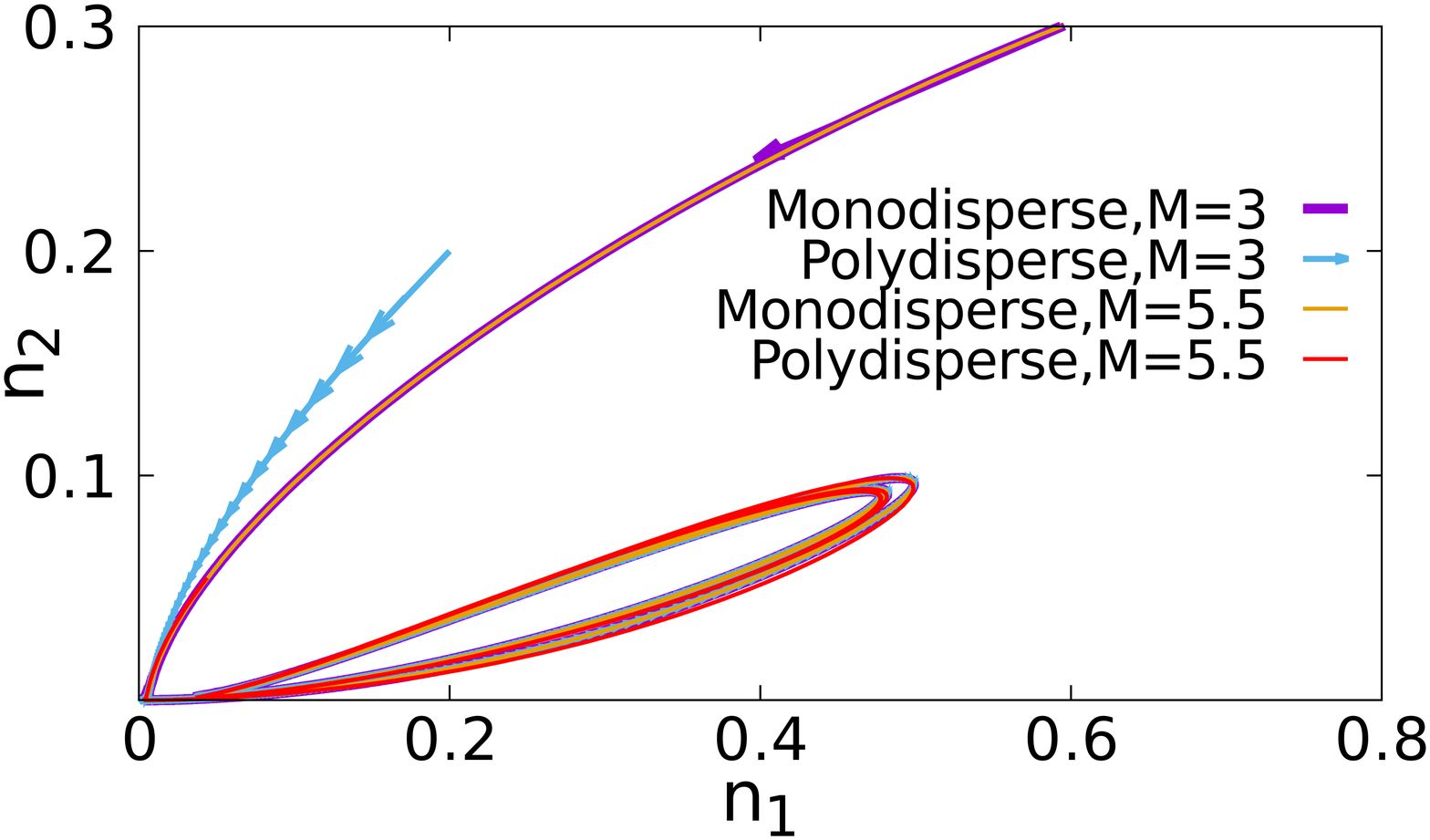}
\end{center}
\caption{Limit cycle for the steady-state oscillations in terms of $n_1(t)$ and $n_2(t)$ for $a=0.95$ and
$\lambda =0.005$. For the total mass density $M=5.5$ the initial conditions are mono-disperse, $n_k=M
\delta_{1,k}$ and stepwise, \eqref{inicon}.  For $M=3$  the initial and current values of $n_1(t)$ and
$n_2(t)$ are re-scaled accordingly. The relaxation to the unique limit cycle is clearly visible. }
\label{pic:limit_cycle}
\end{figure}

In Fig.~\ref{pic:Osillations_integral} we show oscillating solutions for $N(t)$ for $0.9< a \leq 1$. The new
feature observed in Fig.~\ref{pic:Osillations_integral} is the emergence of stationary oscillations. All
cluster concentration $n_k(t)$ perform these stable oscillations; the form of the oscillations depends on the
cluster size and the amplitude decreases with the increasing size, see Fig.~\ref{pic:Oscil_095098}. Figure \ref{pic:limit_cycle}, demonstrates that the system reaches a limit cycle~\footnote{The definition of a limit cycle in the system~\eqref{eq:Model} and \eqref{eq:Cija} with homogeneous kernels $K_{i,j}$ and $F_{i,j}$ has some subtleties discussed in the SM.} which does not depend on the total mass or initial conditions.

Our results indicate the existence of a critical value of $\lambda_c(a)$ such  that for $\lambda < \lambda_c(a)$ the steady-state solution is no longer stable and instead the system approaches to a limit cycle. Although for $a<0.9$ we have observed only damped oscillations, we believe that stationary oscillations would emerge for all $a \geq 1/2$ but the required values of $\lambda $ are too small. When $\lambda$ is small, a huge number of equations is needed to achieve a requested precision. For example, for the simulations presented in Fig.~\ref{pic:Damping_osillations} already $200,000$ equations have been used. For systems with $\lambda < \lambda_c$ for $a<0.9$, one needs to solve $N_{\rm eq}> 10^6$ nonlinear
ODEs which is a formidable task even when fast numerical methods are applied (see the SM).

Our major observations may be summarized as follows: 
\begin{enumerate}
\item When $a<1/2$, there exists  a single stable fixed point for all values of $\lambda$; the steady state distribution of cluster sizes $n_k$ corresponds to this fixed point.
\item When $1/2 < a\leq 1$ and $\lambda > \lambda_c(a)$, the system has a single stable fixed point; it may be a stable
focus, resulting in dumped oscillations.
\item When $1/2 < a\leq 1$ and $\lambda < \lambda_c(a)$, the system has an attractive limit cycle. 
\end{enumerate}
In the $1/2<a \leq 1 $ range, the above assertions are conjectural and require further verification.

To find the steady-state cluster size distribution we set $dn_1/dt =0$
and $dn_k/dt=0$ in Eqs.~\eqref{eq:Model} and solve the resulting infinite system of algebraic equations. Introducing the generating functions ${\cal C}_{\pm a}(z) = \sum_{k\geq 1} k^{\pm a} n_k z^k$, we transform \eqref{eq:Model} into
\beq
\lb{eq:gengen} {\cal C}_a(z)\,{\cal C}_{-a} (z) &+& (1+\lambda)zn_1(M_a + M_{-a} ) \\
&=& (1+\lambda)\left( M_a {\cal C}_{-a} (z)   - M_{-a} {\cal C}_{a} (z) \right). \nn \eeq
where ${\cal C}_{\pm a}(1) = M_{\pm a}$. Setting $z=1$ in Eq.~\eqref{eq:gengen} yields

\be \lb{eq:MaMa} M_a M_{-a} = \frac{1+\lambda}{1+2 \lambda} n_1 (M_a + M_{-a} ). \ee
To analyze the tail of the size distribution, i.e. $n_k$ for $k \gg 1$, we exploit the methods described e.g.
in \cite{Krapivsky,ColmPaulJCP2012} in the context of similar problems. Recalling that when $a=0$ the tail is $n_k  \simeq \lambda \pi^{-1/2} k^{-3/2}e^{-\lambda^2 k}$ for $k \gg 1$,  see \cite{PNAS}, suggests that $n_k \simeq C
k^{-\tau} e^{-\omega k }$ for $k \gg 1$, with some constants $C$, $\tau$ and $\omega$. Expanding the generating functions ${\cal C}_{\pm a} (z)$ near $z^{\prime}
\to 1-0$, where $z^{\prime} = z/z_0$ and $z_0= e^{\omega}$, we get

\be \lb{eq:Gen_f_exp} {\cal C}_{\pm a} (z) = {\cal C}_{\pm a} (z_0) + C \Gamma(1 \pm a -\tau) (1-
z^{\prime})^{\tau \mp a-1}. \ee
Here $\Gamma(x)$ is the gamma function and we assume that ${\cal C}_{a} (z_0)  =\sum_{k \geq 1} k^a n_k z_0^k
< \infty$ exists for given $a$ and $\tau$. If we substitute the above ${\cal C}_{\pm a} (z)$ into
Eq.~\eqref{eq:gengen} we obtain terms with different powers of the factor $(1-z^{\prime})$. To satisfy this
equation  we equate to zero all these terms separately. In this way we obtain equations for the zero-order
terms of $(1-z^{\prime})$, and the terms of the order of $(1-z^{\prime})^{\tau \pm a -1}$. Combining these
equations with Eq.~\eqref{eq:MaMa} we find $\tau=3/2$ and $\omega \simeq \lambda^2$, and finally the
amplitude $C=M\lambda \pi^{-1/2}$ (see SM for details). Thus the tail of the cluster size distribution reads

\be \lb{eq:disfun2} n_k \simeq  \lambda \pi^{-1/2} M k^{-3/2}  e^{-\lambda^2 k}   \qq {\rm for}   \qq k \gg
1. \ee
In the above analysis we assume that ${\cal C}_a (z_0)$ exists. This is a consistent assumption when $a<1/2$,
but fails for $a \geq 1/2$ (see the SM) thereby manifesting a qualitative change in the system dynamics,
which we indeed observe in simulations.

To conclude, we investigated numerically and analytically a system of particles undergoing aggregation
and collision-controlled shattering (the complete fragmentation into monomers). We considered spatially
homogeneous well-mixed systems characterized by the aggregation kernel $K_{i,j} = (i/j)^a + (j/i)^a$ and shattering kernel $F_{i,j} =\lambda K_{i,j}$.  For $a<1/2$, we obtained an analytical solution for the steady-state cluster  size distribution and confirmed numerically the relaxation of the size distribution to this steady-state form. For $a \geq  1/2$, the temporal behavior drastically depends on the shattering constant $\lambda$: When $\lambda > \lambda_c(a)$ the system relaxes to a steady-state through damped oscillations, while for $\lambda < \lambda_c(a)$ oscillations become stationary and persist forever (Figs.~\ref{pic:Osillations_integral}--\ref{pic:limit_cycle}).

Using the language of dynamical systems our observations can be reformulated as  follows: (i)~for $a<1/2$ the
governing system of ODEs possesses a single stable fixed point for all values of $\lambda$, (ii)~for $1/2
\leq a \leq 1$, the system has a single stable fixed point (which may be a stable focus) when $\lambda \geq
\lambda_c(a)$, and (iii)~for $1/2 \leq a \leq 1$ and $\lambda < \lambda_c(a)$ the system possesses a stable
limit cycle.

Limit cycles may arise already for two coupled ODEs \cite{Strogatz}. Still, the emergence of stable oscillations in a {\em closed} system comprising an infinite number  of species and undergoing aggregating and shattering is striking. To the
best of our knowledge this phenomenon has not been previously observed and a relaxation towards the steady
state was believed to be the only possible scenario.

\medskip
The work was supported by the Russian Science Foundation, grant 14-11-00806.

\bibliography{oscillations}

\begin{widetext}

\newpage

\section*{\large Supplemental Material: Oscillations in aggregation-shattering processes}

\section*{\large Numerical versus analytical solutions for the constant kernels}

Figure \ref{pic:cmp_analyt} illustrates the application of the fast-integration method for the case of constant aggregation and shattering kernels ($a=0$). The numerical solution approaches to the steady-state solution which is known analytically \cite{PNAS}:
\begin{equation*}
\label{nk-sol} n_k=\frac{N}{\sqrt{4\pi}} \left(1+\lambda\right)\left[\frac{2n_1}{(1+\lambda)N}\right]^k\,
\frac{\Gamma(k-\frac{1}{2})}{\Gamma(k+1)} \,.
\end{equation*}
Here $n_1=\lambda/(1+\lambda)$ is the stationary density of monomers and $N=2\lambda/(1+2\lambda)$ is the stationary density of clusters. The above distribution refers to the case of unit mass density, $M=1$, the general case is obtained by multiplying the above densities by $M$. For $\lambda \ll 1 $ and $k \gg 1$, the above exact distribution simplifies to
\begin{equation}
\label{nk-asymp-simple} 
n_k=\frac{\lambda}{\sqrt{\pi}}\,e^{-\lambda^2 k}\,k^{-3/2} \,.
\end{equation}
Figure \ref{pic:cmp_analyt} demonstrates the high accuracy of the numerical method and the intuitively obvious feature that  the smaller the parameter $\lambda$ the longer it takes for the system to reach the steady state (when $\lambda=0$, the steady state is never reached).

\begin{figure}[h!]
\begin{center}
\includegraphics[scale=0.6, page=1]{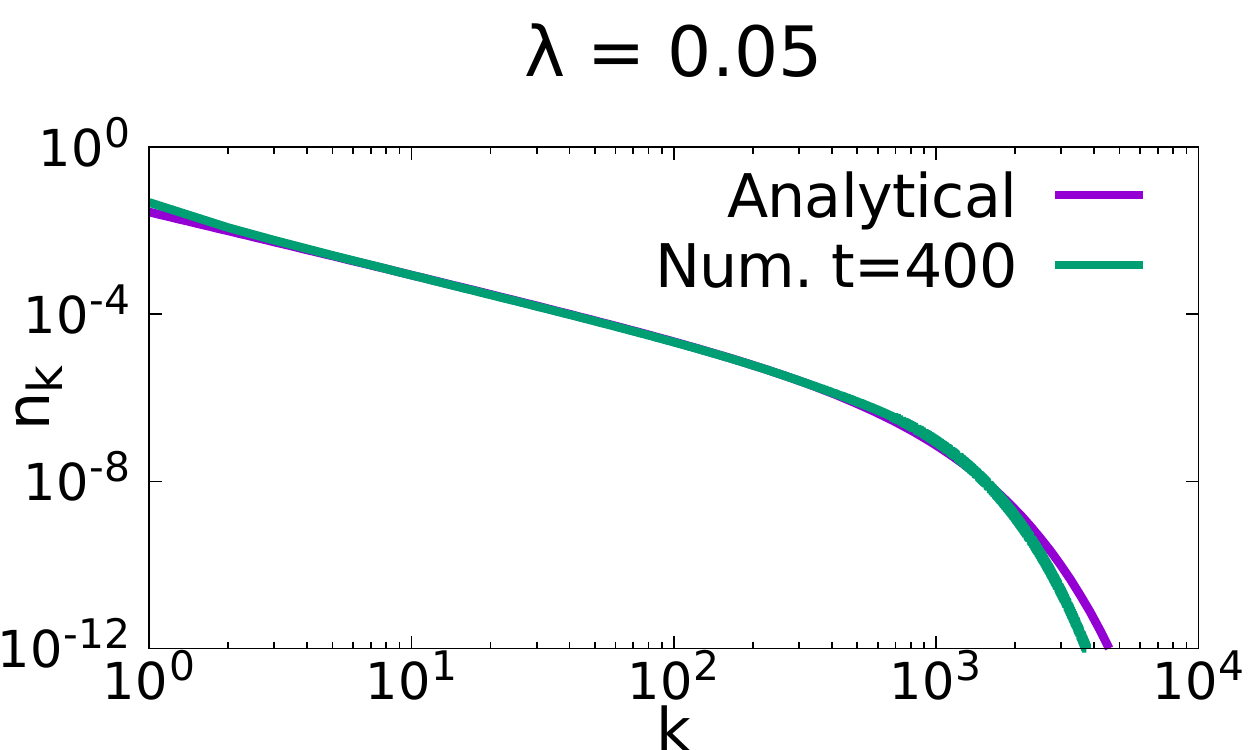} \qq \qq \qq
\includegraphics[scale=0.6, page=3]{cmp_analyt.pdf}
\caption{Comparison of the numerical solution with the analytical steady-state solution~\cite{PNAS} for the
constant kernel $K_{i,j} = 1$ and monodisperse initial conditions, $n_k(0)=M\delta_{1,k}$ with $M=1$, for
$\lambda=0.05$ (left panel) and $\lambda=0.02$ (right panel).} \label{pic:cmp_analyt}
\end{center}
\end{figure}

\section*{\large Limit Cycles of kinetic  equations with homogeneous kernels}

One must be careful while talking about limit cycles for kinetic equations with homogeneous kernels. By definition, a limit cycle is an {\em isolated} closed trajectory; this means that its neighboring trajectories are not closed --- they spiral either towards or away from the limit cycle~\cite{Strogatz}. After this definition, we are usually told that limit cycles can only occur in
nonlinear systems; in a linear system exhibiting oscillations closed trajectories are neighbored by other
closed trajectories. We also learn that a {\em stable} limit cycle is one which attracts all neighboring
trajectories. A system with a stable limit cycle can exhibit self-sustained oscillations~\cite{Strogatz}.

Consider a dynamical system that may be written as
\begin{equation}
\label{AF:gen} 
\frac{dn_k}{dt} = F_k({\bf n}), \quad k\geq 1
\end{equation}
where ${\bf n} = (n_1, n_2, n_3, \ldots)$ and $F_k({\bf n})$ is given in our case by Eqs.~(3) and (4) of the
main text. It is important to note that the reaction terms $F_k({\bf n})$ are strictly quadratic polynomials
for all $k$, and this fact alone leads to the conclusion that limit cycles in the dynamical system
\eqref{AF:gen} are impossible. Indeed, Eqs.~\eqref{AF:gen} are invariant under the transformation
\begin{equation}
\label{trans} t\to T/M, \quad n_k\to MN_k
\end{equation}
namely after this transformation Eqs.~\eqref{AF:gen} become
\begin{equation}
\label{AF:trans} \frac{dN_k}{dT} = F_k({\bf N})
\end{equation}
with the same functions $F_k$. Therefore if the dynamical system \eqref{AF:gen} possesses a limit cycle, we
can slightly perturb it by choosing $M=1+\epsilon$ with $|\epsilon|\ll 1$ and obtain another limit cycle
implying that a closed trajectory is not isolated and hence it is not a limit cycle. (More generally,
the dynamical system \eqref{AF:gen} in which $F_k({\bf n})$ for all $k$ are homogeneous polynomials of any
degree does not have limit cycles.)

We now recall that our dynamical system actually admits an integral of motion, namely the mass density is
conserved:
\begin{equation}
\label{mass} 
\sum_{j\geq 1} jn_j = 1.
\end{equation}
In Eq.~\eqref{mass} we have set the mass density to unity; if the mass density is equal to $M$, we can make
the transformation \eqref{trans} and then the mass density will be equal to unity.

The original dynamical system \eqref{AF:gen} was considered in the (infinite-dimensional) quadrant
\begin{equation}
\label{quadrant} 
\mathbb{R}_+^{\infty}=\{(n_1, n_2, n_3, \ldots)| n_1\geq 0, n_2\geq 0, n_3\geq 0, \ldots\}.
\end{equation}
But it is more appropriate to reduce \eqref{AF:gen} to the phase space which is an intersection of
\eqref{mass} and \eqref{quadrant}. Plugging $n_1 = 1 - \sum_{j\geq 2} jn_j$ into \eqref{AF:gen} with $k\geq
2$ we obtain
\begin{equation}
\label{AF:red} \frac{dn_k}{dt} = G_k({\bf n}), \quad k\geq 2.
\end{equation}
The functions $G_k({\bf n})$ are still quadratic polynomials, but not strictly quadratic. For example, the
term $n_1n_2$ turns into $n_2-\sum_{j\geq 2} jn_j n_2$.

The dynamical system \eqref{AF:red} is defined on
\begin{equation}
\label{simplex} 
\left\{(n_2, n_3, \ldots)| n_2\geq 0, n_3\geq 0, \ldots; \sum_{j \geq 2} jn_j\leq 1\right\}
\end{equation}
which is the (infinite-dimensional) simplex. This dynamical system may admit genuine limit cycles. Hence all
dynamical systems \eqref{AF:gen} and \eqref{quadrant} with different masses $M$ are equivalent to the generic
one, given by Eqs.~\eqref{AF:red} and \eqref{simplex}. In other words, systems with different masses $M$ may
be mapped on each other by simple re-scaling. Most of our simulations  have been done for the stepwise
initial distribution of the cluster sizes,

\begin{equation}
\label{IC:10} 
n_j(0) =
\begin{cases}
0.1 & 1\leq j\leq 10\\
0    & j>10,
\end{cases}
\end{equation}
with the total mass $M=5.5$. To illustrate that our limit cycle is unique (up to the numerical precision) we
also consider the total mass of $M=3$ and mono-disperse initial conditions $n_k(0)=M\delta_{1,k}$. In Fig.~6
of the main text it is demonstrated that the closed trajectories of the $n_1(t)-n_2(t)$ plane coincide for
different masses and initial conditions after the appropriate re-scaling. This proves numerically the
existence of true limit cycle in the system of interest.

\section*{\large Analytical approach for the stationary distribution}

To find the steady-state distribution of the aggregates sizes, one needs to put $\dot{n}_1=\dot{n}_k=0$ into
the left-hand side of Eqs.~(3) and (4) of the main text and solve the following infinite system of algebraic
equations:
\begin{eqnarray}
\label{eq:Model_stead}
&&n_1 \sum\limits_{i=1}^{\infty} K_{1i} n_i - \frac{\lambda}{2} \sum_{i = 2}^{\infty} \sum_{j=2}^{\infty} (i+j) K_{i,j}
n_i~ n_j -
\lambda n_1 \sum_{j=2}^{\infty} j K_{1,j} n_j =0 \\
&& \frac{1}{2} \sum\limits_{i=1}^{k-1} K_{i,k-i} n_i n_{k-i} - (1 + \lambda) n_k \sum\limits_{i=1}^{\infty}
K_{k,i} n_i =0,  \qquad \qquad k \geq 2. \nonumber
\end{eqnarray}
We will apply the method of generating functions that has proved its efficiency for similar
problems~\cite{Krapivsky,Leyvraz,ColmPaulJCP2012}. Namely, we introduce the generating functions ${\cal
C}_{\pm a}(z) $ and moments $M_{\pm a}$:

\be \lb{eq:genfun} {\cal C}_{\pm a} (z)= \sum_{k=1}^{\infty} k^{\pm a} n_k z^k \qq  
M_{\pm a} = \sum_{k=1}^{\infty} k^{\pm a} n_k,
\ee
Multiplying \eqref{eq:Model_stead} by $z^k$ and summing over all $k\geq 1$ we arrive at

\be 
\lb{gengen} {\cal C}_a(z){\cal C}_{-a} (z) +(1+\lambda)zn_1(M_a + M_{-a} ) - (1+\lambda)\left(M_a
{\cal C}_{-a} (z) + M_{-a} {\cal C}_{a} (z) \right)=0. \ee
Specializing \eqref{gengen} to $z=1$ and taking into account that  ${\cal C}_{\pm a} (1) = M_{\pm a}$ we obtain 

\be 
\lb{MaMa} M_a M_{-a} = \frac{1+\lambda}{1+2 \lambda} n_1 (M_a + M_{-a} ). \ee
The tail of the size distribution can be extracted from the asymptotic behavior of the generation functions ${\cal C}_a
(z)$. We consider separately the cases of $a<1/2 $ and $a > 1/2$.

\emph{Kernels with $a<1/2$. } The tail \eqref{nk-asymp-simple} arising in the context of the model with constant kernel, $a=0$, in the case when additionally $\lambda\ll 1$,  suggests that generally steady-state distribution may have a similar tail,

\be \lb{eq:st_st} 
n_k  \simeq C k^{-\tau} e^{-\omega k } \qq \qq {\rm for} \qq  \qq k \gg 1,  \ee
for kernels with $a>0$. The amplitudes $C$ and $\omega$ and the exponent $\tau$ are yet unknown functions of $\lambda$ and $a$. The generation functions may be expanded near $z^{\prime} \to 1-0$, where $z^{\prime} = z/z_0$ and $z_0= e^{\omega}$. One seeks the expansions in the form \cite{Krapivsky,ColmPaulJCP2012}

\beq \lb{Gen_f_exp} {\cal C}_{a} (z) &=& {\cal C}_{a} (z_0) + C \Gamma(1+a -\tau) (1-
z^{\prime})^{\tau-a-1} \\
{\cal C}_{-a} (z) &=& {\cal C}_{-a} (z_0) + C \Gamma(1-a -\tau) (1- z^{\prime})^{\tau+a-1},  \eeq
where $\Gamma(x)$ is the gamma function and we assume that ${\cal C}_{a} (z_0)  =\sum_{k \geq 1} k^a n_k
z_0^k < \infty$ exists for given $a$ and $\tau$. If we substitute the above ${\cal C}_{\pm a} (z)$ into
Eq.~\eqref{eq:gengen} we obtain terms with different powers of the factor $(1-z^{\prime})$. To satisfy this
equation  we equate to zero all these terms separately. Then the zero-order terms of $(1-z^{\prime})$ yield

\be \lb{eq:zero} 
{\cal C}_{a} (z_0){\cal C}_{-a} (z_0) -(1+\lambda) \left( M_{a}{\cal C}_{-a} (z_0) +M_{-a}
{\cal C}_{a} (z_0) \right) + (1+\lambda)z_0 n_1 (M_a + M_{-a} )=0. \ee
Similarly, the terms of the order $(1-z^{\prime})^{\tau \pm a -1}$ imply the relations

\be \lb{eq:pma} {\cal C}_{\mp a} (z_0)  \Gamma(1 \pm a -\tau) -(1+\lambda) M_{\mp a}   \Gamma(1 \pm a
-\tau)=0. \ee
Finally, the rest of the terms should satisfy

\be \lb{eq:fineq} C^2 \Gamma(1 + a -\tau) \Gamma(1 - a -\tau) (1-z^{\prime})^{2\tau -2 } -(1+\lambda)z_0 n_1
(M_a + M_{-a} ) (1-z^{\prime}) =0 . \ee
This equation is consistent when

\be \lb{eq:tau} 2\tau -2 = 1 \qq \qq {\rm or } \qq \qq \tau =\frac{3}{2}\, .\ee
The exponent $\tau$ is therefore universal (independent on $a$ and $\lambda$). Now we substitute the relations

\be \lb{eq:MaCa} {\cal C}_{\mp a} (z_0) =(1+\lambda) M_{\mp a},  \ee
which follow from Eq.~\eqref{eq:pma} into Eq.~\eqref{eq:zero} to yield 

\be \lb{eq:MaMa1} M_a M_{-a} = \frac{z_0 }{1+  \lambda} n_1 (M_a + M_{-a} ). \ee
From Eqs.~\eqref{eq:MaMa1} and \eqref{eq:MaMa} then follows,

\be \lb{eq:z0} z_0 =e^{\omega} = \frac{(1+\lambda)^2}{(1 +2 \lambda)}.\ee
The ansatz \eqref{eq:st_st} is expected to work only when $\lambda\ll 1$. In this limit \eqref{eq:z0} gives 
$\omega \simeq \lambda^2 - 2 \lambda^3 + \ldots \simeq \lambda^2$, so the amplitude $\omega$ is independent on $a$.  
Thus the tail of the cluster size distribution reads 

\be \lb{eq:tail} n_k \simeq \frac{C}{k^{3/2} } e^{-\lambda^2 k}  \qq \qq {\rm for} \qq  \qq k \gg 1.\ee
An order-of-magnitude estimate for the constant $C$ may be done as follows. We assume that the distribution
\eqref{eq:tail}, which holds true for $k \gg 1$ may be also used for $k \sim 1$, so that

\be \sum_{k=1}^{\infty} k n_k \simeq \int_1^{\infty} \frac{C}{ k^{1/2}} e^{-\lambda^2 k} dk  \simeq \frac{C
\sqrt{\pi}}{\lambda} =1 , \ee
that is, $C \simeq  \lambda/\sqrt{\pi} \sim \lambda $.

\emph{Kernels with $a > 1/2$. } Applying the same  analysis  as above for $a \geq 1/2$, one arrives at
Eqs.~\eqref{eq:zero}--\eqref{eq:fineq}, which however do not lead to consistent results. Indeed, from
Eq.~\eqref{eq:fineq} it follows that $\tau =3/2$, but ${\cal C}_a (z_0)$ does not exist for $a \geq 1/2$, so
that Eq.~\eqref{eq:pma} may not be satisfied to cancel the terms corresponding to the factor
$(1-z^{\prime})^{\tau + a -1}$.

Although the above approach fails to make consistent asymptotic estimates for $a \geq 1/2$, our results for
$a<1/2$ and the results of Ref.~\cite{connaughton2016universality} for a similar system motivate as to
exploit a hypothesis,  that for $a \geq 1/2$, the distribution of cluster size has the following form for $k
\gg 1$: $n_k \simeq C k^{-3/2}e^{-\lambda^{\beta} k}$; it will be used  below for the further analysis.

\section*{\large Truncating an infinite system of equations by a finite number of equations}

The standard problem of numerical solution of Smoluchowski equations is how to approximate an infinite system
of equations by a finite one. When fragmentation is lacking, as in common Smoluchowski equations, the average
size of aggregates infinitely grows which imposes a principle time limit for the modeled processes. Contrary,
in the case of interest, the fragmentation of aggregates precludes the formation of very large clusters even
for infinitely long time. Therefore the number of equations may be finite. Moreover, using the results for
the steady-state distribution, one can estimate the number of equations needed to describe the system with a
given degree of accuracy. Below we show, how the solutions of a formally infinite system (3) of the main text
may be adequately represented by these of a finite system.

Using Eqs.~(3) of the main text, we write for the concentration $n_k(t)$ for $2 \leq k \leq L$:

\begin{equation}
\label{eq:With_tail}
\frac{d n_k}{dt } = \frac{1}{2} \sum\limits_{i=1}^{k-1} K_{i,k-i} n_i n_{k-i} - (1 + \lambda) n_k
\sum\limits_{i=1}^{L} K_{k,i} n_i - \left\lbrace (1 + \lambda) n_k \sum\limits_{i=L+1}^{\infty} K_{k, i} n_i
\right\rbrace.
\end{equation}
Taking into account that for $a \leq  1$ and $k < L < i$
$$
K_{i,k} = \left(\frac{i}{k} \right)^a + \left(\frac{k}{i} \right)^a \leq  i^a + k^a < i + i = 2i
$$
and applying the steady-state distribution,
$$
n_k \simeq  C k^{-\frac{3}{2}} e^{-\lambda^{\beta} k}
$$
we estimate the factor in the curled bracket in \eqref{eq:With_tail} as
\begin{align*}
  \sum_{i=L+1}^{\infty} K_{i,k} n_i  < 2
\sum_{i=L+1}^{\infty} i n_i
 \sim \int_{L}^{\infty} x C x^{-\frac{3}{2}} e^{-\lambda^{\beta} x } dx  \sim C
L^{\frac{1}{2}} \frac{\text{Erfc}(\sqrt{\lambda^{\beta} L})}{\sqrt{\lambda^{\beta} L}} .
\end{align*}
If the quantity $\lambda^{\beta} L$ is large, one can make a further simplification, using the asymptotic
relation $\text{Erfc}(x) \sim e^{-x^2}/x$, which yields,

\be \lb{eq:estN}
 \sum_{i=L+1}^{\infty} K_{i,k} n_i <
 C \frac{e^{-\lambda^{\beta} L}}{\lambda^{\beta} L^{\frac{1}{2}}}  < \varepsilon.
\ee
 Hence, if we choose the number of equations $ L=N_{\rm eq}(\lambda)$ such that the above expression is smaller
than $\varepsilon \ll 1$ we can safely skip the term in the curled brackets in Eq.~\eqref{eq:With_tail} to
obtain:
\begin{equation}\notag
\frac{d n_k}{dt } = \frac{1}{2} \sum\limits_{i=1}^{k-1} K_{i,k-i} n_i n_{k-i} - (1 + \lambda) n_k
\sum\limits_{i=1}^{N_{\rm eq}} K_{ki} n_i,
\end{equation}
that is, the solution of an infinite system may be approximated with any desired accuracy by the solution of
a finite system with the appropriately chosen number of equations. In practice, we started with the number of
equations estimated from Eq.~\eqref{eq:estN} for $\beta =2$ and $C\sim \lambda$, as for the steady-state
distribution for $a<1/2$ and checked, whether the simulation results keep unchanged (within the machine
precision) when the number of equations increases. For the most of studied systems the appropriate number of
equations was about $N_{\rm eq} =150,000$.

\end{widetext}

\end{document}